\documentclass[twocolumn]{aastex62}
\usepackage{longtable}
\usepackage{amsmath}
\usepackage{subfigure}
\DeclareMathOperator{\sech}{sech}

\graphicspath{{./}{figures/}}
\begin{document}

\title{Exploring the Impact of Ejecta Velocity Profile on Kilonova Evolution: Diversity of the Kilonova Lightcurves}

\author[0000-0002-9852-2469]{Donggeun Tak}
\affiliation{SNU Astronomy Research Center, Seoul National University, Seoul 08826, Republic of Korea, \href{donggeun.tak@gmail.com}{donggeun.tak@gmail.com}}
\affiliation{Korea Astronomy and Space Science Institute, Daejeon 34055, Republic of Korea, \href{uhm@kasi.re.kr}{uhm@kasi.re.kr}}
\author{Z. Lucas Uhm}
\altaffiliation{Corresponding author: Z. Lucas Uhm}
\affiliation{Korea Astronomy and Space Science Institute, Daejeon 34055, Republic of Korea, \href{uhm@kasi.re.kr}{uhm@kasi.re.kr}}

\author[0000-0002-8094-6108]{James H. Gillanders}
\affiliation{Astrophysics sub-Department, Department of Physics, University of Oxford, Keble Road, Oxford, OX1 3RH, UK}

\begin{abstract}
A kilonova is a short-lived explosive event in the universe, resulting from the merger of two compact objects. Despite its importance as a primary source of heavy elements through r-process nucleosynthesis, its nature is not well understood, due to its rarity. In this work, we introduce a model that determines the density of a radially-stratified relativistic ejecta. We apply the model to kilonova ejecta and explore several hypothesized velocity profiles as a function of the merger's ejection time. These velocity profiles result in diverse density profiles of the ejecta, for which we conduct radiative transfer simulations using \textsc{tardis} with the solar r-process composition. Consequently, we investigate the impact of the ejecta velocity profile on the resulting lightcurve and spectral evolution through the line transitions of heavy elements. The change in the rate at which these elements accumulate in the line-forming region leaves its imprint on the kilonova lightcurve at specific wavelengths, causing the lightcurves to decay at different rates. Furthermore, in several profiles, plateau-like behaviors (slow and/or flat decline) are also observed. In conclusion, this work proposes potential scenarios of the kilonova evolution due to the ejecta velocity profile.
\end{abstract}

%% Keywords should appear after the \end{abstract} command.
%% See the online documentation for the full list of available subject
%% keywords and the rules for their use.

\section{Introduction}

Mergers of two neutrons stars (NSs) or a neutron star and a stellar-mass black hole (BH) eject a small fraction of matter ($10^{-3}-10^{-1} M_\odot$) into space at (sub-relativistic) speeds $\lesssim 0.35 \, c$ \citep{Kasen2017, Metzger2019}. The neutron-rich ejecta material rapidly decompresses from high densities, and the r-process nucleosynthesis forges heavy elements in the ejecta. Subsequently, these elements decay radioactively and become a long-term heat source for the ejecta lasting weeks, thus powering the kilonova emission. The kilonova emission is thought to be attributed to more than one ejecta component \citep[e.g.,][]{Kasen2017, Perego2017, Siegel2017, Tanaka2017, Radice2018}. The dynamical ejecta is expected to have an ejecta speed of $v \sim 0.2 - 0.3 c$ with neutron-rich material (electron fraction $Y_e \lesssim 0.1$), which is ejected over a timescale of few milliseconds. On the other hand, the disk winds have $v \sim 0.1 c$ with a higher electron fraction ($Y_e \sim 0.2-0.3$) on longer timescales of several hundreds of milliseconds.

In the study of supernovae, the density profile of the ejecta envelope and its composition are key ingredients to determine the evolution of a supernova at various wavelengths. For Type Ia supernovae, a quasi-exponential profile known as the W7 model \citep{Nomoto1984, Branch1985, Thielemann1986} has found extensive application since the model effectively reproduces the distribution of elements with high accuracy \citep[e.g.,][]{Jeffery1992, Mazzali1993, Dutta2021}. While a power-law profile is widely used for Type II supernova studies \citep[e.g.,][]{Blinnikov2000, Dessart2006}, various other profiles have also been adopted \citep[e.g., a combination of distinct inner and outer profiles;][]{Nagy2014, Szalai2016}.

In the case of kilonovae, a consensus regarding the ejecta density profile has not yet been reached. To explain the observational data of kilonovae, various functional forms of the density profile, heavily relying on our understanding of supernovae, have been adopted; e.g., a constant density profile with a sharp cutoff \citep{Darbha2020}, a broken power-law \citep{Kasen2017}, a power-law \citep{Tanaka2017, Watson2019, Gillanders2022}, and multi-dimensional anisotropic profiles \cite[e.g.,][]{Bulla2019, Kawaguchi2020, Breschi2021, Heinzel2021, Korobkin2021, Wollaeger2021}.

Deriving the kilonova ejecta profile would be more complicated, as the mass ejection from the merger and the properties of the ejected flow depend on the binary parameters and the NS equation of state (EOS). Such details have been vigorously studied with hydrodynamic and numerical simulations \citep[e.g.,][]{Hotokezaka2013, Tanaka2013, Kyutoku2015, Siegel2017, Radice2018, Kruger2020, Nedora2022}. For example, \cite{Kyutoku2015} simulated the merger of black hole-neutron star binaries and showed that depending on key parameters such as the EOS, mass ratio, and dimensionless spin parameter a range of complicated ejecta profiles can be obtained. On the other hand, \cite{Radice2018} performed numerical simulations on the merger of binary neutron stars, incorporating the EOS and binary parameters. From the simulations, they extracted dynamical ejecta properties and then derived the density profiles of ejecta as a function of the entropy and electron fraction of materials.

The complexity behind the merging process can lead to diversity in the ejecta velocity profile. As the ejecta propagates, the parts of ejecta with high values of velocity gradient experience significant spread-out or dilution in radial direction and become the regions of low density. In other words, the gradient or stratification in the ejecta velocity profile can give rise to various forms of the ejecta density profile.

In view of no consensus regarding the kilonova ejecta profile, in this work, we hypothesize several velocity profiles as a function of the merger's ejection time. We introduce an ejecta model that calculates the ejecta density profile for those velocity profiles with various stratification. We further investigate how the ejecta velocity profiles leave their imprints on the kilonova lightcurves by performing a radiative-transfer simulation. The ejecta model is presented in Section~\ref{sec:model}. The method of the radiative-transfer simulation is presented in Section~\ref{sec:method}. The result and the discussion are followed in Section~\ref{sec:result} and Section~\ref{sec:conclusion}, respectively.

\section{Ejecta model}\label{sec:model}
\begin{figure*}
  \centering
  \includegraphics[width=0.9\textwidth]{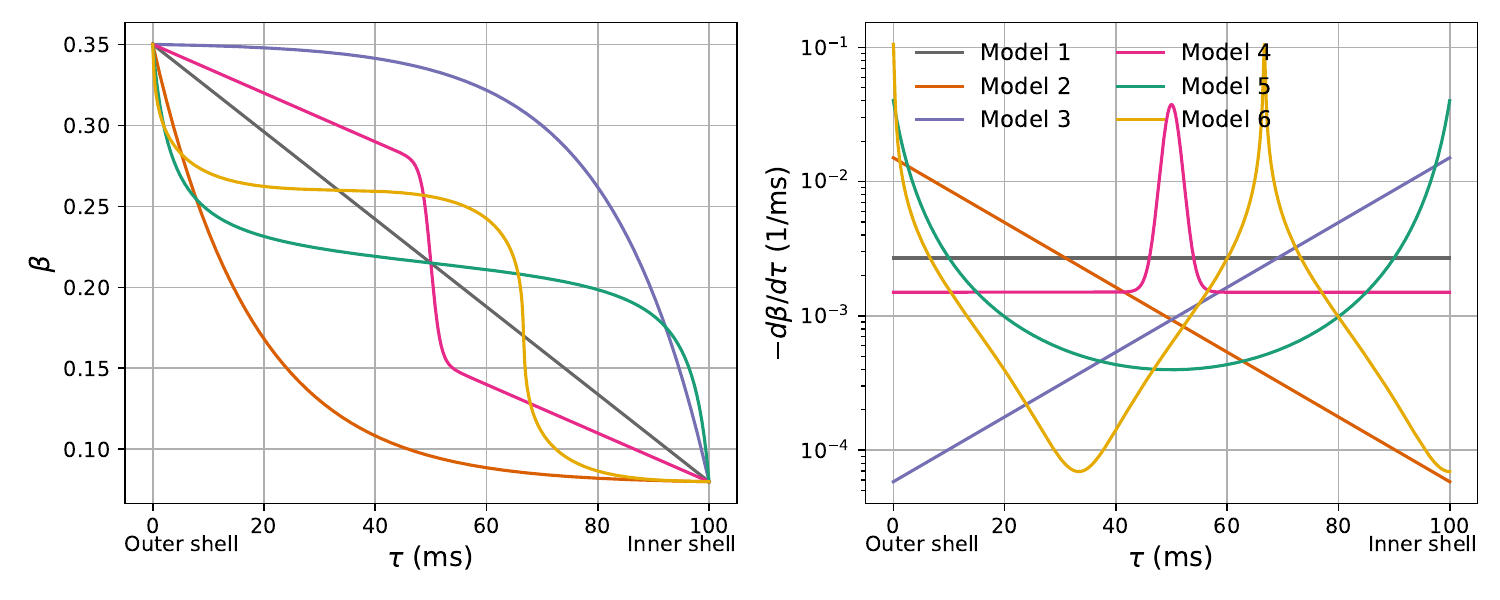}
  \caption{Velocity and its derivative profile of the six ejection scenarios (Models 1 to 6) as a function of the ejection time $\tau$. The left panel shows the velocity profile, and the right panel shows its derivative profile. We assume that the ejection continues up to 100 milliseconds with the outermost shell velocity of 0.35 c and the minimum velocity of 0.08 c.}
  \label{fig:velocity}
\end{figure*}

\begin{figure*}
  \centering
  \includegraphics[width=0.9\textwidth]{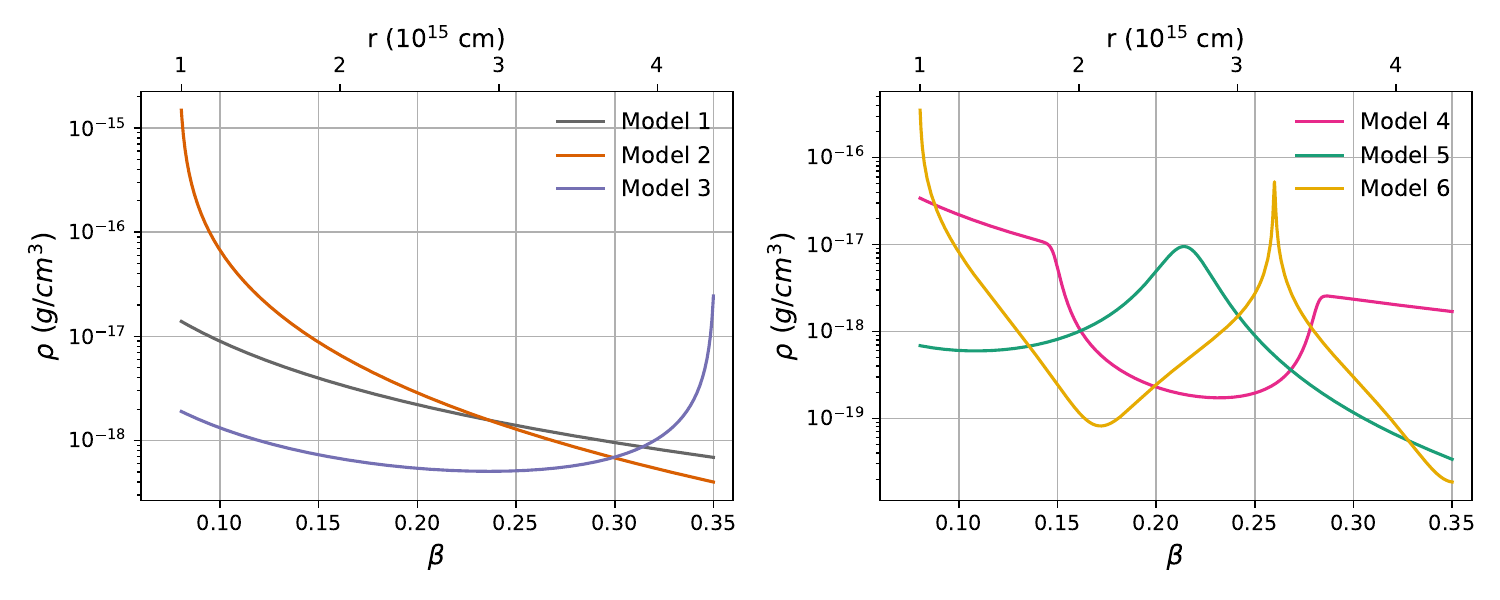}
  \caption{Ejecta density profiles of the six scenarios at 4.8 days after the explosion. The left panel displays the density profiles from the simple cases (Models 1--3), while the right panel show those from the complicated scenarios (Models 4--6). The color scheme labeling the six Models is the same as in Figure~\ref{fig:velocity}.}
  \label{fig:density}
\end{figure*}

We adopt here the ejecta model given in \cite{uhm2011}, which describes a radially-stratified relativistic ejecta of gamma-ray bursts (GRBs). This model makes use of a Lagrangian description and views the ejecta as a sequence of infinitesimal shells that are ejected from the merger. Each shell is prescribed an ejection time $\tau$ as its Lagrangian coordinate. After collisions between adjacent shells, every shell can stationarily coast with its velocity $v(\tau)$ such that any two adjacent shells no longer collide with each other.\footnote{Assuming that the mass ejection occurs at distances of hundreds of kilometers from the merger with sub-relativistic speeds, the shell collisions would cease roughly at the order of seconds. Therefore, this freely-coasting condition may be safely assumed at seconds after the mass ejection. A more accurate timescale for such condition should be derived from hydrodynamic simulations of shell collisions.} Therefore, it is legitimate to consider a non-increasing function of $v(\tau)$: i.e., $dv/d\tau\leq0$. The relativistic continuity equation for a mass flow in spherical polar coordinates can be written as 
\begin{equation}
\nabla_\alpha (\rho u^\alpha) = \frac{1}{c}\frac{\partial}{\partial t}\bigg\vert_r (\rho\Gamma c) + \frac{1}{r^2}\frac{\partial}{\partial r}\bigg\vert_t(r^2\rho\Gamma v) = 0,
\end{equation}
where $\rho$ is the mass density measured in the rest frame of ejecta, $c$ is the speed of light, $t$ is the time in laboratory frame, $r$ is the radius measured from the merger, $u^\alpha = \Gamma (c, v, 0, 0)$, $\Gamma = (1 - \beta^2)^{-1/2}$, and $\beta = v/c$. Note that the radius $r$ of a $\tau$-shell at time $t$ is given by
\begin{equation}\label{eq:radius}
r = v(\tau)(t - \tau),
\end{equation}
where $t \geq \tau$, $0 \leq \tau \leq \tau_d$, and $\tau_d$ is the duration of the ejection time.
By solving the continuity equation, \cite{uhm2011} provides an exact solution of the ejecta density $\rho$ for each Lagrangian shell $\tau$ at radius $r$,
\begin{equation}\label{eq:density}
\rho(\tau, r) = \frac{\dot{M}}{4\pi r^2 v \Gamma } \left[1-r\frac{v'}{v^2} \right]^{-1},
\end{equation}
where $v'(\tau) \equiv dv/d\tau$ and $\dot{M}(\tau)$ represents the mass ejection rate from the merger. In the three-dimensional expansion, the ejecta cross-sectional area is proportional to $r^2$, which gives the density dilution with $r^{-2}$. The term with $v'$ shows the contribution arising from radial spread-out due to the velocity stratification.

Once the merger shapes its mass ejection rate $\dot{M}(\tau)$ and velocity profile $v(\tau)$ as a function of the ejection time $\tau$, Equation~\ref{eq:radius} gives the radius of every $\tau$-shell and Equation~\ref{eq:density} determines the density of all these $\tau$-shells. Therefore, Equation~\ref{eq:density} offers a model to fully describe the density profile of the kilonova ejecta and its evolution over time. This model is very general such that $\dot{M}(\tau)$ and $v(\tau)$ can be arbitrary functions of $\tau$ as long as $v' \leq 0$. When $t \gg \tau_d$ and $v' \neq 0$, Equation~\ref{eq:radius} and Equation~\ref{eq:density} display the nature of homologous expansion, $r=vt$ and $\rho \propto t^{-3}$, respectively. It is worth mentioning that the density profile, $\rho \propto v^{-3}t^{-3}$, employed in the modeling of AT2017gfo \citep[e.g., ][]{Watson2019, Gillanders2022}, can be achieved by setting $\beta = \sech(\tau/\tau_0+\theta_0)$, where $\tau_0$ and $\theta_0$ are arbitrary constants.

We introduce six velocity-profile models, where the outermost shell velocity is fixed to $\beta_{\text{outer}} = 0.35$, and inner shells have a velocity of $\beta(\tau) = \beta_{\text{outer}} + \int\beta'(\tau)d\tau$, where $\beta' = d\beta/d\tau \leq 0$. We assume $\tau_d = 100$ ms since the mass ejection from the merger can persist for several tenths of milliseconds \citep{Kasen2017}. The minimum velocity is set to 0.08 c when $\tau = \tau_d$. Figure~\ref{fig:velocity} shows the velocity (left) and derivative (right) profiles of our six models. Models 1 to 3 represent relatively simple cases, where the derivative is either constant or changing monotonically with time. Models 4 to 6 encompass more complex scenarios, which could reflect irregular ejecting activities of the merger. We remark that these models are not based on any prior information or numerical studies regarding the ejecta velocity profile. Instead, we conceive these models to explore various scenarios, including some complicated cases, as several hydrodynamic simulations have shown irregularities in the density profile \citep[e.g.,][]{Kyutoku2015, Radice2018}. 

The ejected shells outside the optically thick photosphere constitute the so-called line-forming (\textit{lf}) region, where photons interact with the expanding ejecta material, resulting in the line-transition signature in the kilonova spectrum \citep[for a schematic illustration, see e.g., Figure 2 in][]{Gillanders2022}. The innermost boundary of the line-forming region is fixed to a radius of $r_{min} = 10^{15}$ cm from the merger. The velocity of the innermost boundary at a given time is obtained by $v_{min}=r_{min}/t$ since $t \gg \tau_d$ for the timescale of days. This implies that, at a given time, only a portion of the velocity profile between the innermost and outermost boundaries constitutes the density profile of the line-forming region. Note that the minimum velocity reduces to about 0.08 c at 4.8 days after the merger time. For each model, the mass ejection rate, $\dot{M}$, in Equation~\ref{eq:density} is assumed to remain constant over the ejection time $\tau$ and is determined by setting the mass within the line-forming region to be $M_{lf} = 2 \times 10^{-4}$ M$_\odot$ at 2.4 days after the explosion. In other words, each model has a distinct value of $\dot{M}$ while maintaining the same mass in $M_{lf}$ at 2.4 days. Note that $M_{ej} = M_{ph} + M_{lf}$, where $M_{ej}$ is the total ejecta mass, and $M_{ph}$ is the mass beneath the photosphere. 

Figure~\ref{fig:density} shows the density profile as a function of its velocity and radius at 4.8 days after the explosion, as a reference. Due to the radial stratification, various density profiles can be obtained, especially in Models 4 to 6 (see right panel of Figure~\ref{fig:density}). To explain in more detail, the relative velocity difference between the inner and outer boundaries of a shell (equivalent to $\beta'$) affects the density dilution within the shell; the higher $|\beta'|$, the higher the dilution. For example, in Model 6, we have an extremely high $|\beta'|$ at $\beta \sim 0.15$ (or $\tau\sim$ 67 ms), leading to an exceedingly low density at that velocity.

\section{Radiative-transfer spectral analysis}\label{sec:method}
We perform the radiative-transfer spectral analysis for the ejecta models, using the well-established code, \textsc{tardis}
\citep[][]{Kerzendorf2014, tardis}.
\textsc{tardis} is a widely-used Monte Carlo radiative-transfer spectral synthesis software for supernova \citep[e.g.,][]{Dutta2021, Williamson2021} and kilonova simulations \citep[e.g.,][]{Smartt2017, Gillanders2022}. {Assuming a spherically-symmetric explosion with an optically thick region beneath the innermost boundary of the line-forming region, \textsc{tardis} computes what the resultant spectrum will be after the primary photospheric spectrum propagates through the homologously-expanding outer regions of ejecta.

We follow the \textsc{tardis} setup implemented by \cite{Gillanders2022}, which successfully reproduced the AT2017gfo spectra during the photospheric epochs. Here we also use the same atomic data set, which, in addition to the default \textsc{tardis} atomic data set,\footnote{The default \textsc{tardis} atomic data is composed of Chianti data for H and He \citep{Zanna2021}, and Kurucz data for the other elements \citep{Kurucz2018}.} contains the extended Kurucz atomic data \citep[Sr$_{\text{I-III}}$, Y$_{\text{I-II}}$, and Zr$_{\text{I-III}}$;][]{Kurucz2018}, the Database on Rare Earths At Mons university \citep[DREAM; 57$\leq$Z$\leq$71;][]{Quinet2020}, and the Pt$_{\text{I-III}}$ and Au$_{\text{I-III}}$ atomic data presented by \cite{Gillanders2021} and \cite{McCann2022}.
In total, we include 1,029,577 transition lines from 49,117 energy levels in elements  of 1$\leq$Z$\leq$92 \citep[for details, see][]{Gillanders2022}. 

Since we aim to explore the effect of the stratified velocity model, we refrain from varying the abundance and instead invoke the solar-system r-process composition here, adopted from \cite{Prantzos2020}.

For the photospheric spectrum, we assume a steady-state photosphere: its temperature $T$ decreases linearly in time, 
\begin{equation}\label{eq:temp}
T(t_d) = 4080-200\, t_d \quad (K),
\end{equation}
where $t_d$ is the time in days since the merger. This formula is approximately derived using the results of \cite{Gillanders2022}, where their model temperatures of AT2017gfo decrease by $\sim 200$ \textit{K} per day, from 2.4 days to 4.4 days after the explosion.\footnote{Note that the estimated temperature of the observed spectrum at 1.4 days in \cite{Gillanders2022} does not follow Equation~\ref{eq:temp}. Since they found that the emission at 1.4 days has a distinct composition to later epochs, it is reasonable to expect the early ejecta to possess a different temperature profile \citep[e.g.,][]{Villar2017}.}
This simple assumption provides the luminosity of the photosphere as a function of time, $L(t) = 4\pi r_{min}^2 \sigma_{\text{R}} T^4$, where $\sigma_{\text{R}}$ is the Stefan-Boltzmann constant. The ejecta temperature is also fixed to the photospheric temperature to avoid complex and inaccurate effects, stemming from the uncertain temperature profile \citep[see][for details]{Gillanders2022}. We acknowledge that these assumptions are not entirely realistic, but they are sufficient to explore the impact of the ejecta velocity profile on kilonova evolution.
% \textcolor{red}{I think we need to add some caveat here that this is the $T$ profile that works for 2017gfo -- not some $T$ profile for kilonovae generally. So what we are really doing is exploring how our best-fitting models for AT2017gfo would vary, assuming different velocity profiles, right? I can add the text, but I need to think about it for a bit.}
% \textcolor{blue}{The sentence beginning with ``We acknowledge'' admits some caveats. Isn't this enough? If not, let's think about how to add some caveats. Also, we can add that this temperature profile provides reasonably good fits for 2017gfo.}
% \dg{maybe we could mention it as an example. You can add them.}

Since the kilonova ejecta is expanding with a relativistic speed during early epochs ($v \gtrsim 0.1 - 0.2 \, c$), we use the full treatment of special relativity in the radiative transfer, including angle aberration, as implemented in \textsc{tardis} by \cite{Vogl2019}. Furthermore, we employ the local thermal equilibrium ({\tt LTE}) approximation to handle ionization and the {\tt dilute-LTE} approach for excitation. To account for line interactions, we adopt the {\tt macro-atom} scheme \citep{Lucy2002, Lucy2003}, which accounts for the effects of fluorescence and multiple internal line transitions.

\section{Result}\label{sec:result}

\begin{figure*}[b!]
    \centering
    \subfigure{\includegraphics[width=0.32\textwidth]{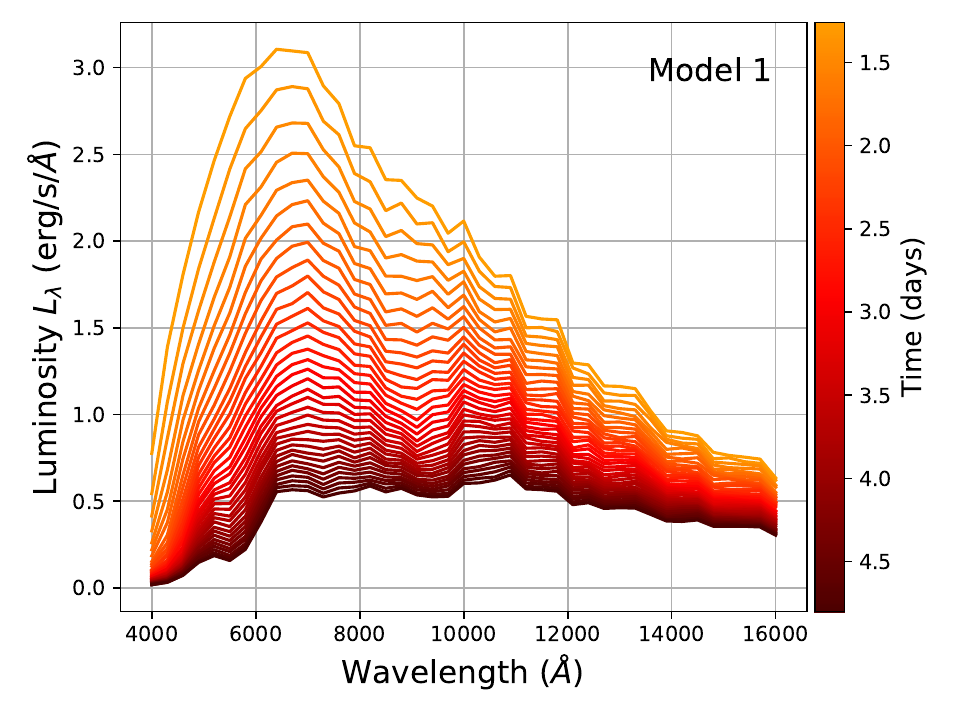}}
    \subfigure{\includegraphics[width=0.32\textwidth]{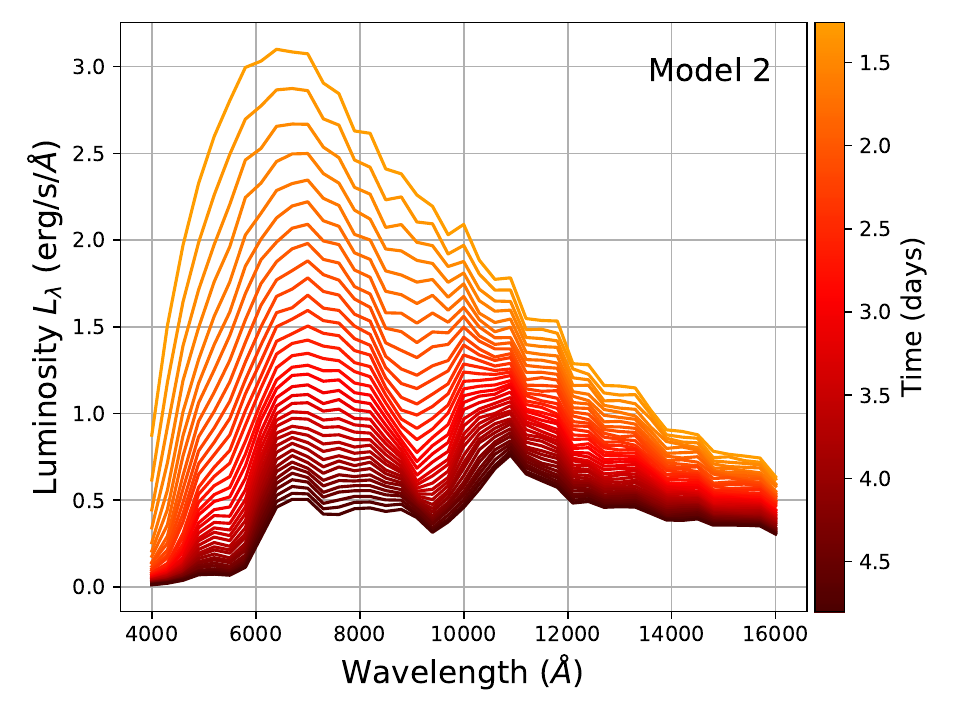}}
    \subfigure{\includegraphics[width=0.32\textwidth]{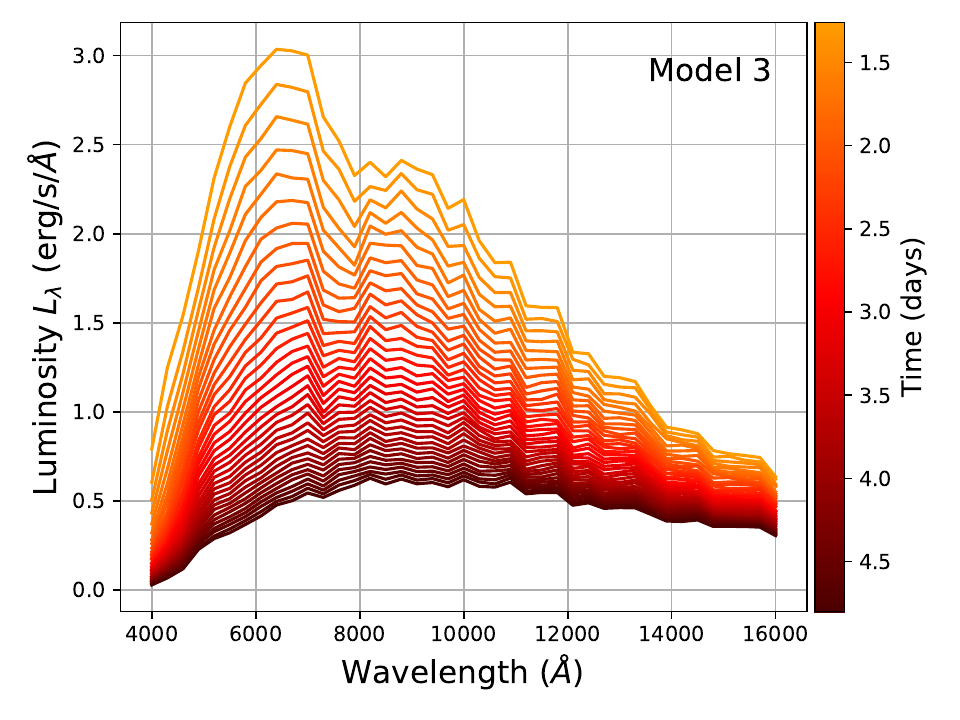}}\\
    \subfigure{\includegraphics[width=0.32\textwidth]{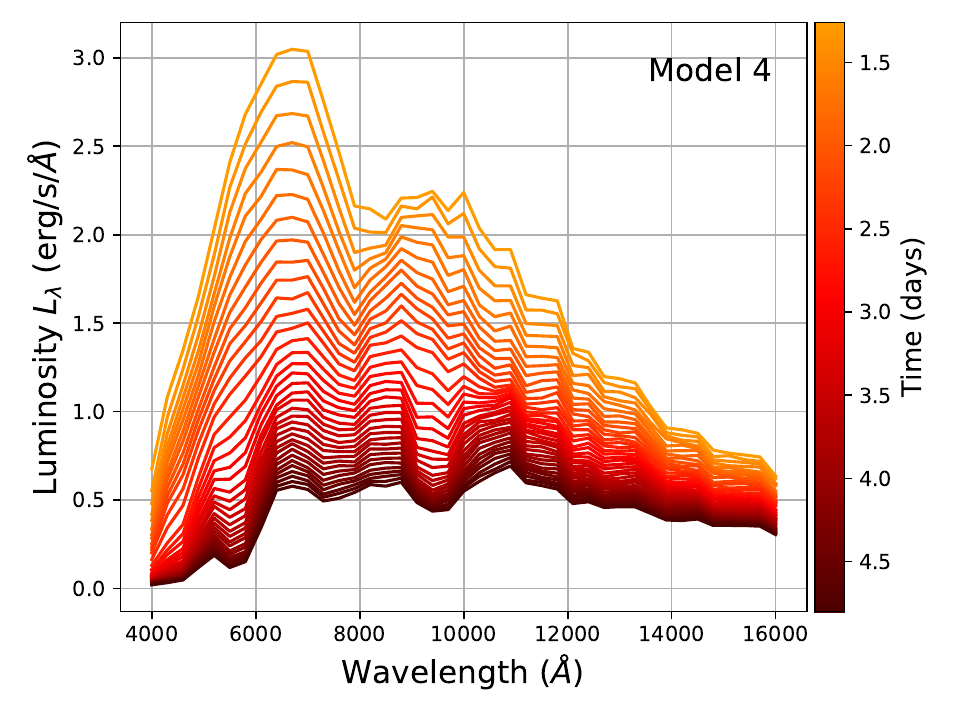}}
    \subfigure{\includegraphics[width=0.32\textwidth]{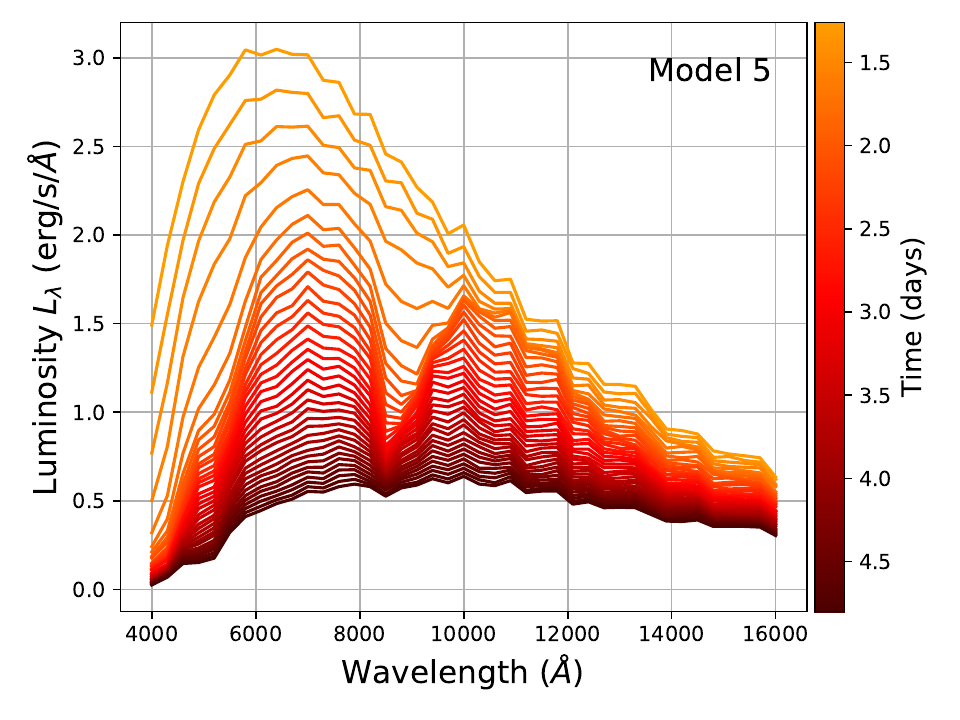}}
    \subfigure{\includegraphics[width=0.32\textwidth]{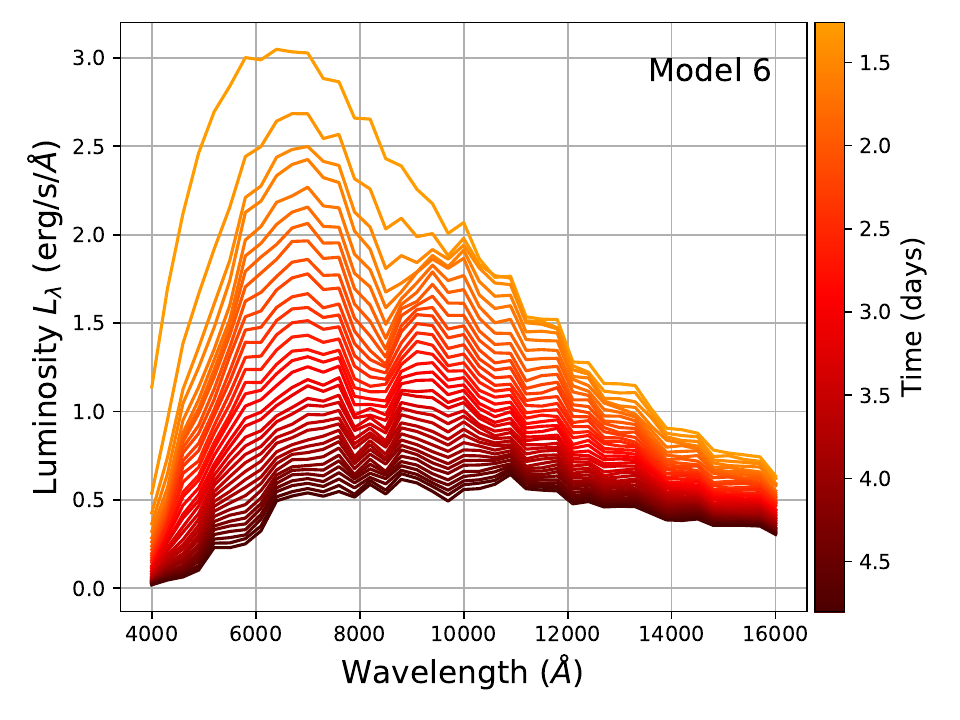}}
    \caption{Evolution of the spectral energy distributions of Models 1 to 6. Each panel shows the spectra of each model at times ranging from 1.4 days to 4.8 days. Deeper colors on the plot represent later times in the evolution.}
    \label{fig:sed_all}
\end{figure*}

\begin{figure*}[b!]
    \centering
    \subfigure{\includegraphics[width=0.32\textwidth]{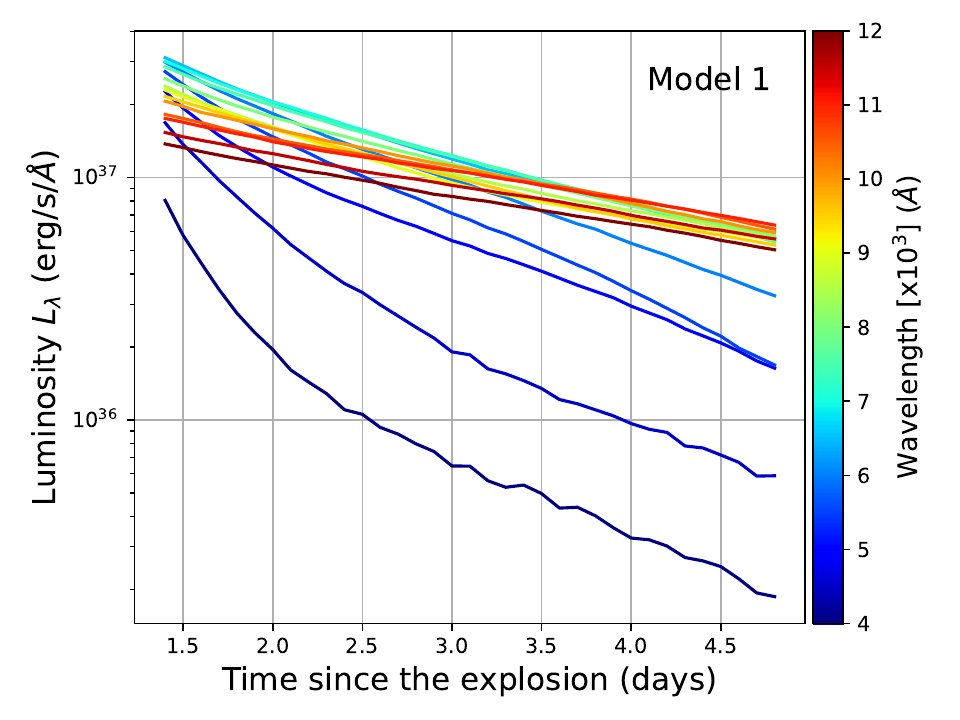}}
    \subfigure{\includegraphics[width=0.32\textwidth]{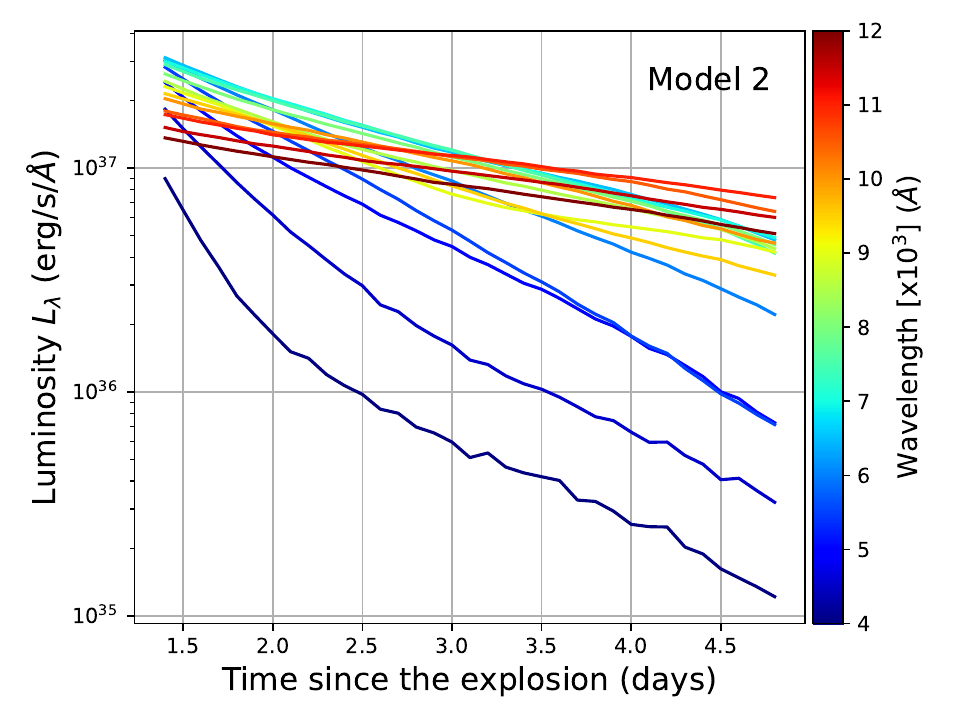}}
    \subfigure{\includegraphics[width=0.32\textwidth]{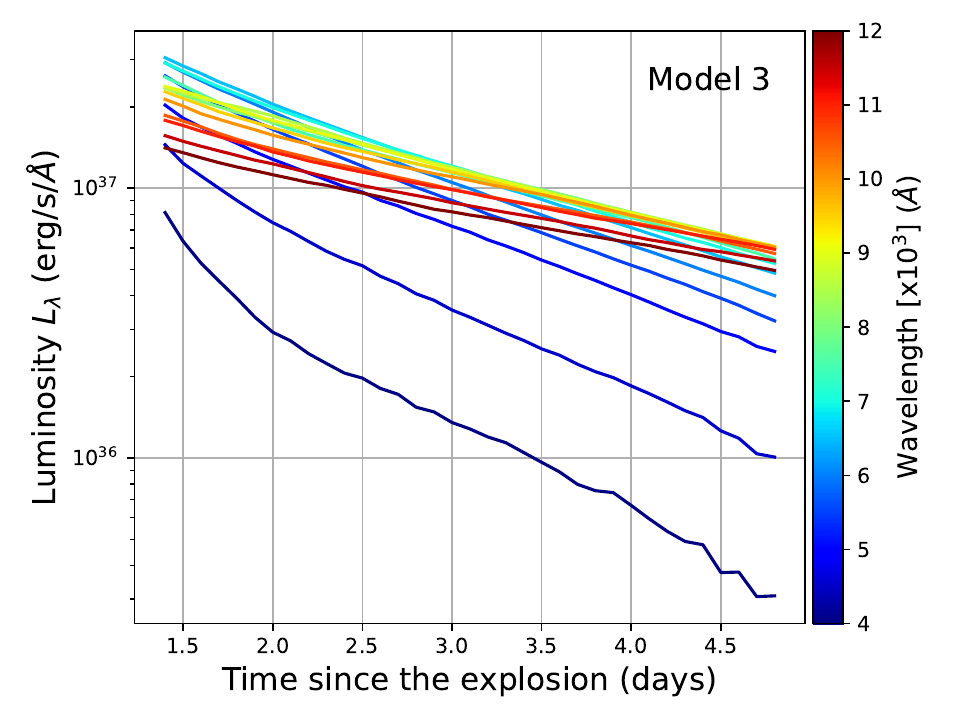}}\\
    \subfigure{\includegraphics[width=0.32\textwidth]{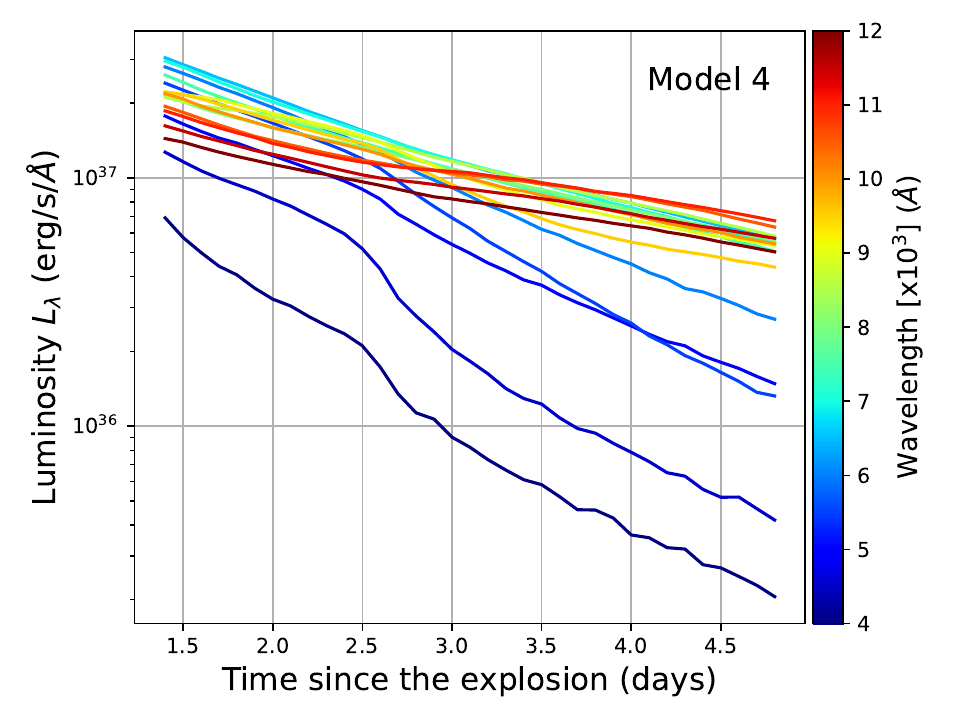}}
    \subfigure{\includegraphics[width=0.32\textwidth]{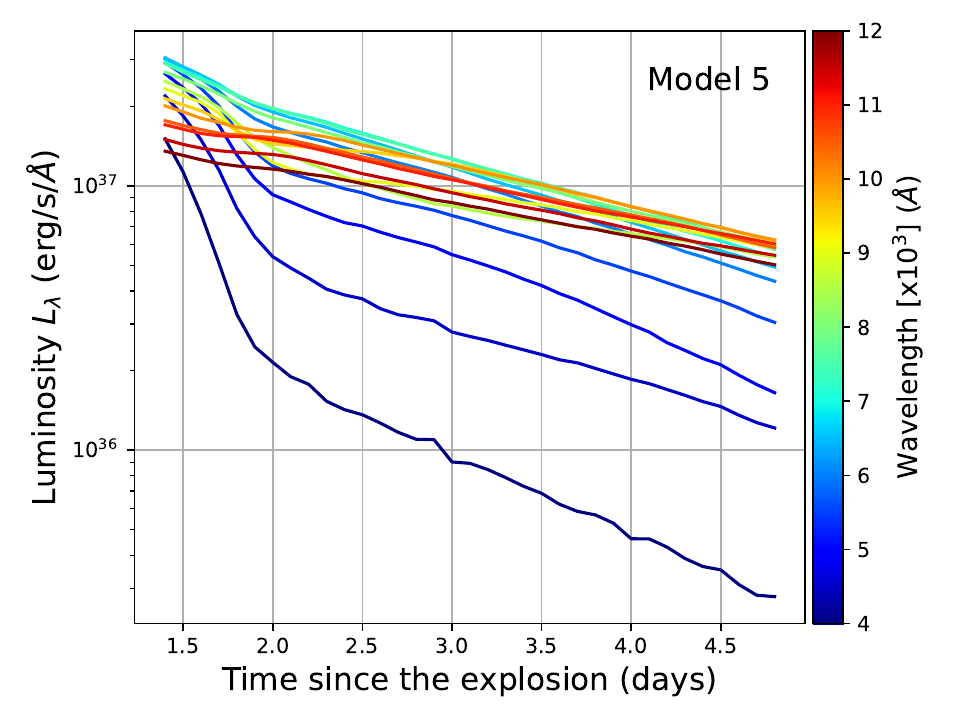}}
    \subfigure{\includegraphics[width=0.32\textwidth]{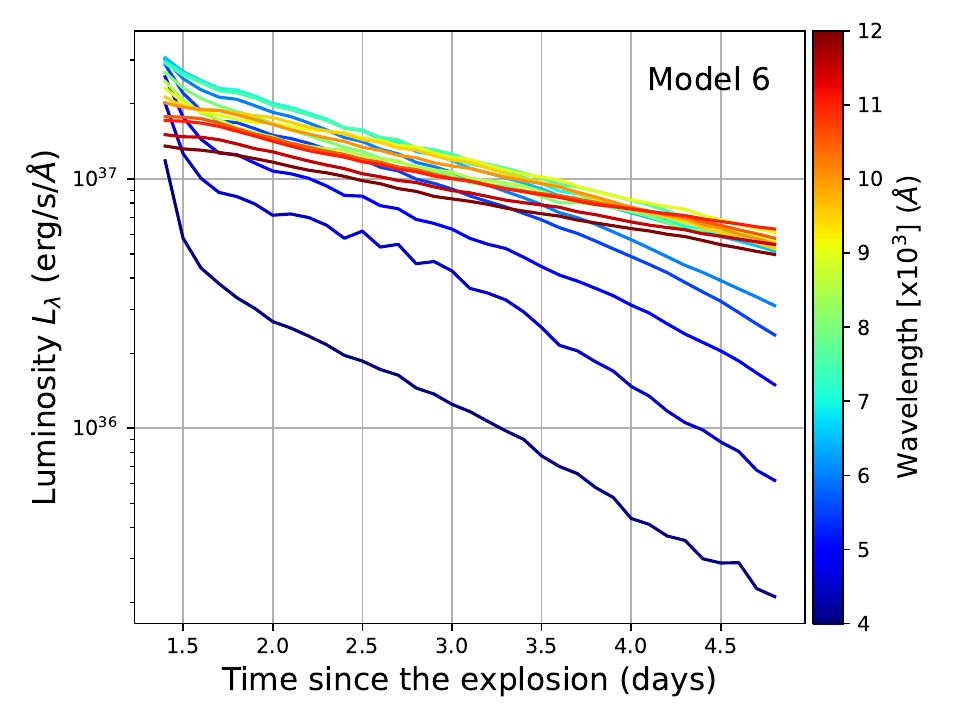}}
    \caption{Luminosity lightcurves of Models 1 to 6 from 1.4 days to 4.8 days. In each panel, the lightcurve is displayed for 16 wavelengths ranging from 4000 $\AA$ to 12000 $\AA$ with the bandwidth of 500 $\AA$. The reddish color indicates higher wavelengths, while the bluish color corresponds to lower wavelengths.}
    \label{fig:lc_all}
\end{figure*}

\begin{figure*}
  \centering
  \includegraphics[width=0.9\textwidth]{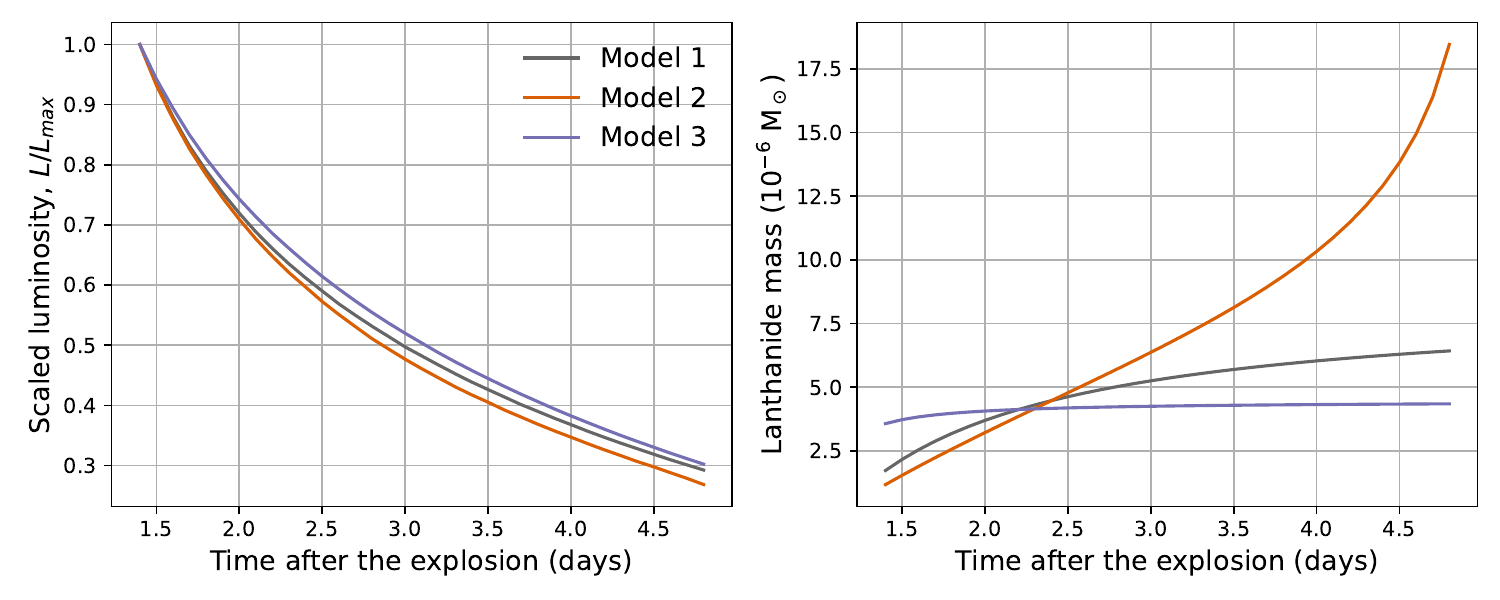}
  \caption{Bolometric luminosity lightcurves and lanthanide mass for simple scenarios. The left panel shows the luminosity lightcurve scaled by its maximum. The right panel shows the total lanthanide mass in the line-forming region as a function of time. Both panels display Models 1 to 3 represented by the colors gray, orange, and purple, respectively.}
  \label{fig:bol_lc}
\end{figure*}

\begin{figure*}
  \centering
    \subfigure{\includegraphics[width=0.32\textwidth]{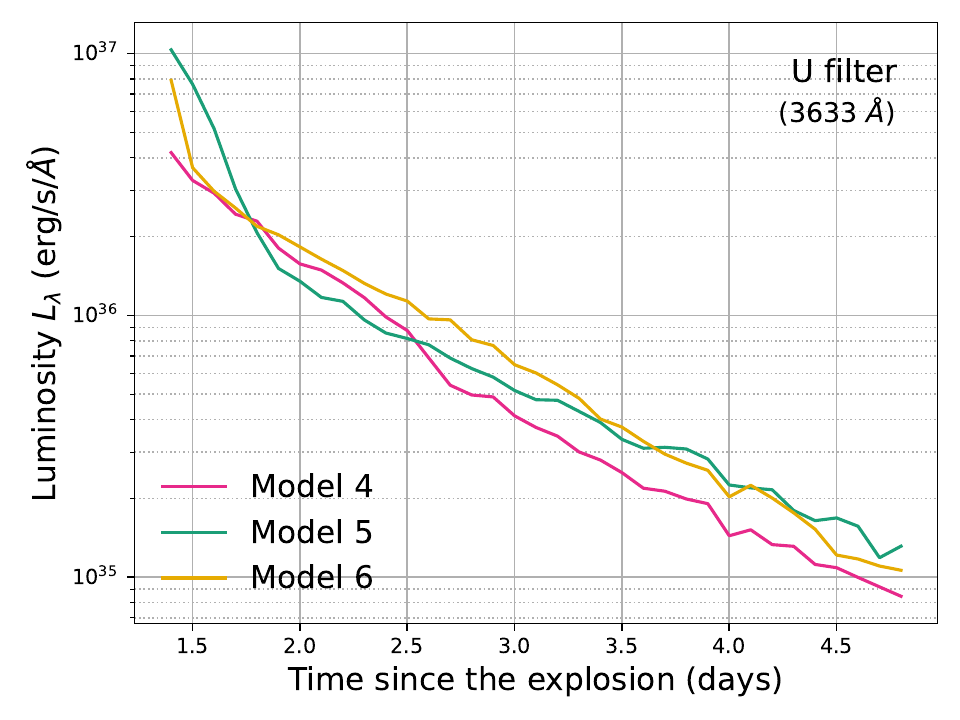}}
    \subfigure{\includegraphics[width=0.32\textwidth]{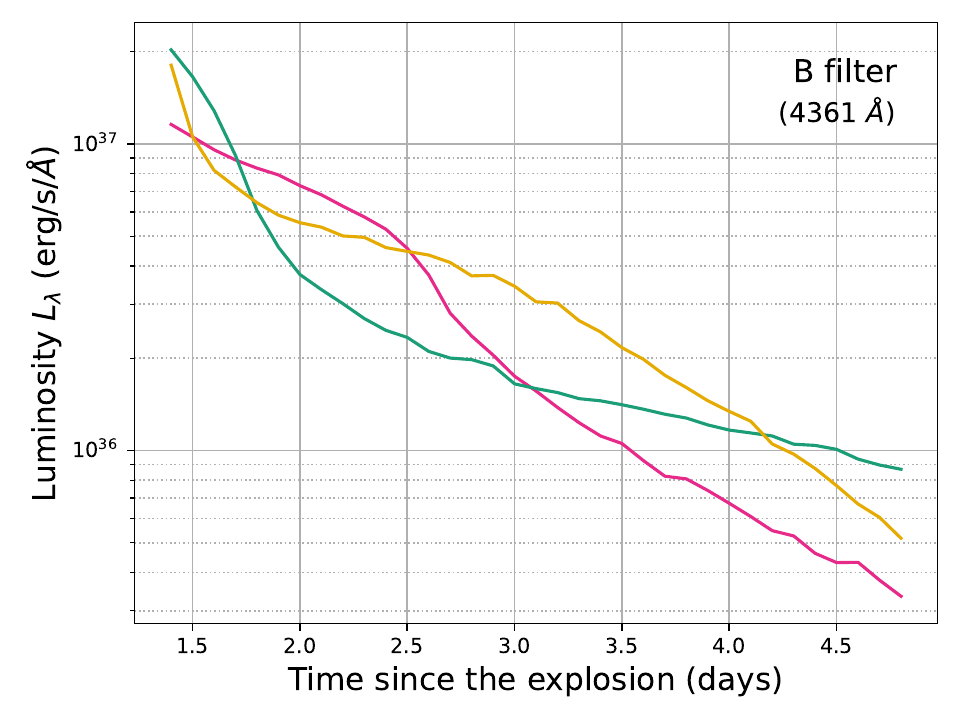}}
    \subfigure{\includegraphics[width=0.32\textwidth]{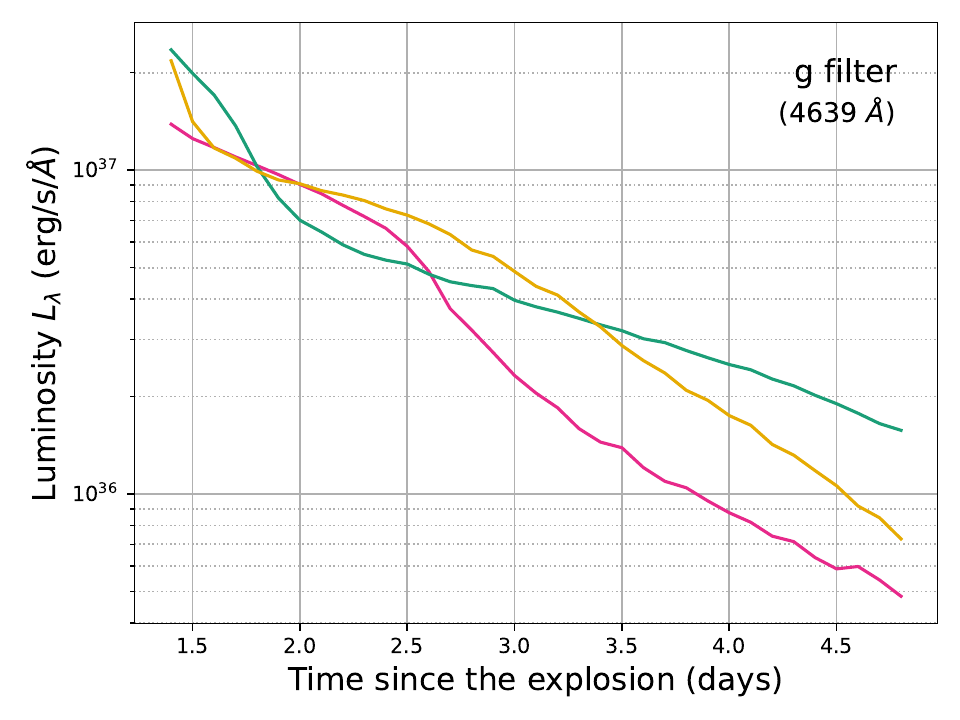}}\\
    \subfigure{\includegraphics[width=0.32\textwidth]{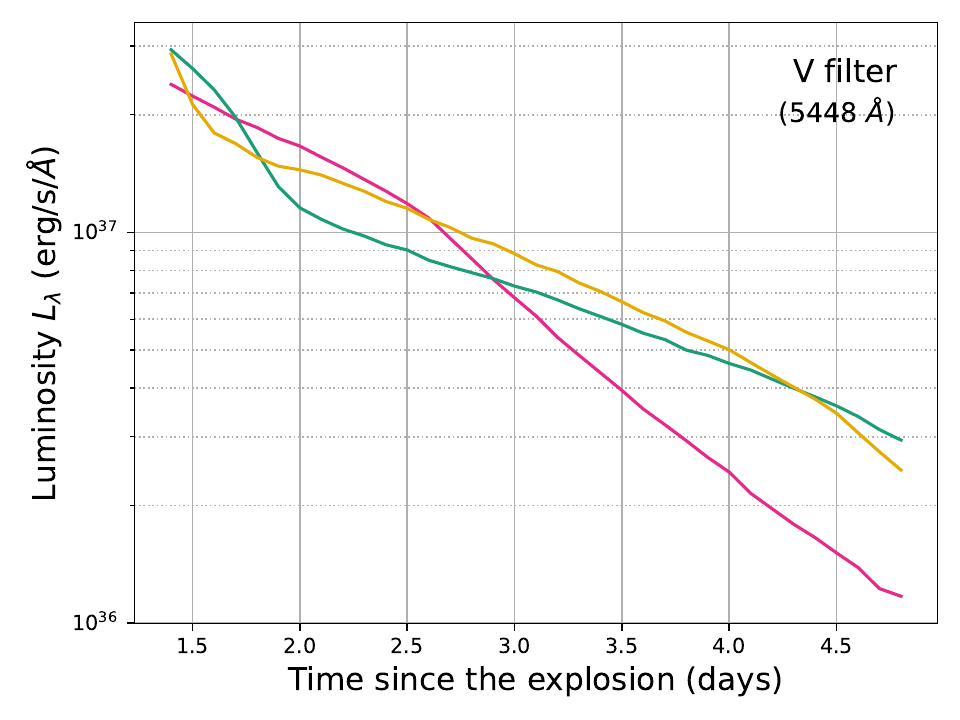}}
    \subfigure{\includegraphics[width=0.32\textwidth]{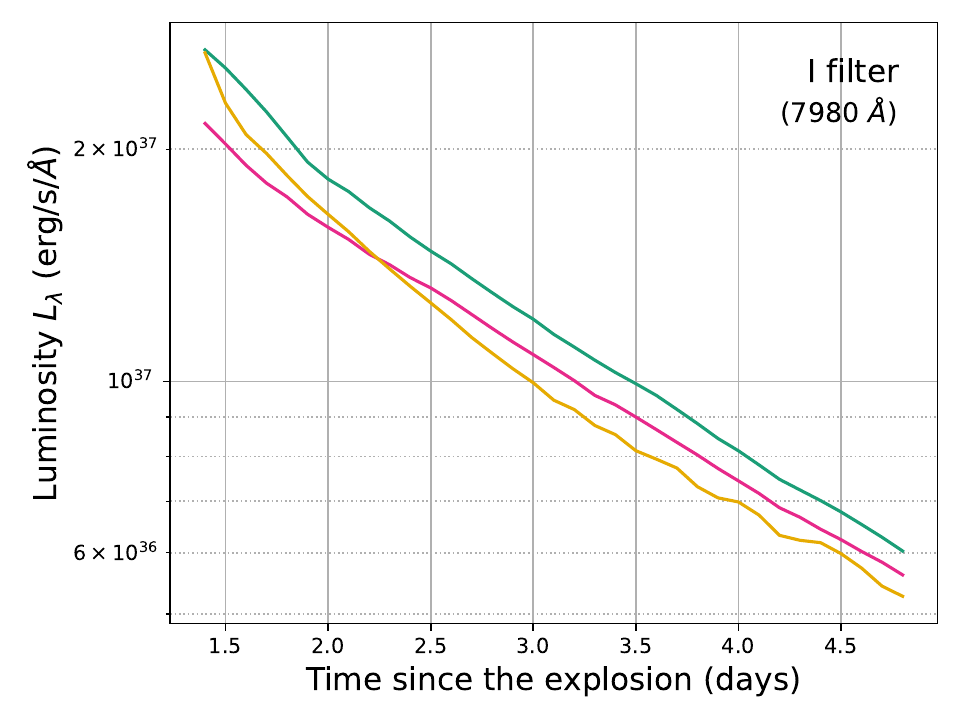}}
    \subfigure{\includegraphics[width=0.32\textwidth]{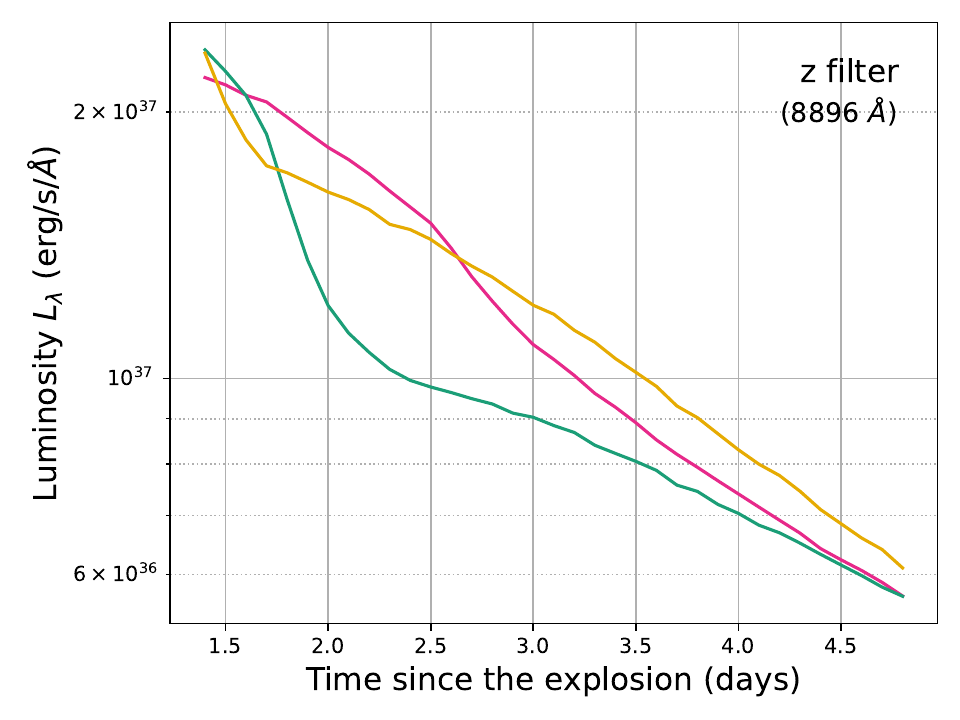}}
    \caption{Luminosity lightcurves of Models 4 to 6 in the six filters: UBgVIz filters. In each panel, the lightcurves from the three models are plotted with pink, green, and yellow, respectively. With the exception of the `I' filter, we can observe a deviation in the decaying slope in all three models.}
    \label{fig:lc_model}
\end{figure*}

We compute the kilonova spectra from 1.4 to 4.8 days after the explosion with a time step of 0.1 days for Models 1 to 6. 

Figure~\ref{fig:sed_all} shows the evolution of the spectral energy distributions (SEDs). An SED with a yellowish color corresponds to the earlier time step, transitioning to a reddish color for later times. Although they originate from identical conditions except for the velocity profile, they undergo different evolutionary paths. Consequently, the SEDs at 4.8 days for the six models are highly distinct. For example, in Model 5 and 6, one can find that the spectra fall rapidly at $\sim$ 8500$\AA$ and form a peak at $\sim$ 10,000$\AA$. This feature is the P-Cygni line profile arising from the Sr\textsubscript{II} near-infrared triplet transitions, as seen in the early epochs of AT2017gfo \citep{Watson2019, Gillanders2022}. Depending on the models and times, this P-Cygni feature can be prominent (or hidden) because its strength relies on the density of Sr\textsubscript{II} at a specific velocity and temperature, as well as the relative line-transition rates of other elements. Further details will be studied in a future work.

Figure~\ref{fig:lc_all} shows the lightcurves at 16 different wavelengths. We generate them between 4000 $\AA$ and 12000 $\AA$ with the bandwidth of 500 $\AA$, resulting in 16 curves. We find that Models 4 to 6 show unusual features in some wavelengths, while the lightcurves from Models 1 to 3 decay smoothly in all wavelengths.

To understand the origin of the distinct evolutions of Models 1-6, we firstly delve into the simple scenarios (Models 1--3) to examine how the concave or convex shape in the velocity profile affects the overall decay rate of the bolometric luminosity. In the wavelengths from 500$\AA$ to 25000$\AA$, the spectral energy distributions of the three models exhibit a gradual decay over time (Figure~\ref{fig:sed_all}). The left panel of Figure~\ref{fig:bol_lc} shows the scaled bolometric luminosity lightcurves for Models 1 to 3. We normalize the bolometric luminosity to its maximum value as we are primarily interested in the relative changes rather than the absolute magnitude of the luminosity. Model 2 shows a relatively steep decay, compared to the others. To investigate the reason behind the difference in decaying rates, we compute the total mass of lanthanide elements (57 $\leq$ Z $\leq$ 71) in the line-forming region, where they contribute significantly to the line-transition opacity at wavelengths below $\sim$ 7000 $\AA$ (the right panel of Figure~\ref{fig:bol_lc}). Note that these three models have similar lanthanide mass at 2.4 days after the explosion because each model is configured to have the same mass in the line-forming region at this epoch (see Section~\ref{sec:model}). The accumulated amount of lanthanide materials of Model 2 increases sharply in time, compared to the others. This leads to an increase in the total absorption rate, resulting in a steeper decay in the lightcurve. 

Compared to the simple scenarios, Models 4 to 6 exhibit diverse features in their lightcurves, attributed to the various density profiles (Figure~\ref{fig:density}). Figure~\ref{fig:lc_model} shows the lightcurves of Models 4 to 6 in UBgVIz filters. In the three models, a significant change in the decay rate is evident in all filters, except for the ``I'' filter. Specifically, Model 4 tends to show a steepening at around 2.5 days, which is particularly prominent in the filters ranging from 4300 to 5500 $\AA$. Conversely, Model 5 demonstrates an early sharp decay followed by a more gradual decline. This break between the two phases occurs at around 2 days. Lastly, Model 6 shows a dynamic evolution in its lightcurve: an initial steep decay, followed by a period of slower decline, and then a subsequent re-steepening. Especially in the ``B" and ``g" filter, the aforementioned evolutions in Models 4 to 6 are conspicuous. We note that the lightcurves from ``u" filter (3596$\AA$) are similar to that of ``U" filter (3633$\AA$). In case of ``r" (6122$\AA$), ``R" (6407$\AA$), and ``i" (7439$\AA$) filters, the luminosity follows a gradual decline over time, resembling the behavior observed in the ``I" filter (lower-middle panel of Figure~\ref{fig:lc_model}).

\section{Discussion and Conclusion}\label{sec:conclusion}

In this work, we explored the impact of ejecta velocity profile on kilonova evolution. For various velocity profiles as a function of the merger's ejection time, we calculated the ejecta density profiles using the ejecta model by \cite{uhm2011} and then computed the resultant lightcurves with the radiative-transfer spectral synthesis software, \textsc{tardis}. The results indicate that the lightcurve of kilonovae can exhibit not only a change in its decaying rate but also a plateau-like behavior (slow and/or flat decline). 

The plateau-like lightcurve is seemingly similar to that of Type II-P supernovae but they have different origins. The plateau in Type II-P supernovae is attributed to a change in the electron-scattering opacity as the ionization and recombination of hydrogen in the supernova's envelope progresses \citep{Grassberg1971}. In the case of kilonovae, the electron scattering contributes much less to the total opacity compared to line transitions by heavy elements \citep{Kasen2013}. In other words, the plateau-like behavior in kilonovae would be observed at specific wavelengths where the heavy elements play an important role in absorbing and re-emitting photons through line transitions. As shown in Figure~\ref{fig:lc_model}, the filters and their corresponding wavelengths that exhibit a notable change in the decay rate are indeed connected to the energy levels at which the dominant line transitions of the heavy elements (e.g., Sr, Ce, Nd, and Sm) mainly take place. As reported in many studies on AT2017gfo \citep[e.g.,][]{Kasen2017, Watson2019, Domoto2021, Gillanders2022}, these elements play a key role in generating absorption and re-emission features in the kilonova spectrum at early epochs.

Generally, as the total mass of r-process elements increases, the lightcurve steepens due to a greater amount of absorption (see Figure~\ref{fig:bol_lc}). This simple picture can help to explain the evolution of Models 4 to 6 as well. In the case of Model 4, between 1.4 days and 2.5 days, the mass of heavy elements within the line-forming region does not show a notable increase due to a well-like dip in the density profile (see the red lines in Figure~\ref{fig:density}). From 2.5 days, the mass accumulation rate increases sharply, resulting in a break in the lightcurve. Conversely, in Model 5, the mass accumulation rate in the line-forming region increases during the early phase up to 2 days and then gradually decreases at later times (see the green lines in Figure~\ref{fig:density}), hence the lightcurve is steep up to 2 days and following which it flattens. The lightcurve of Model 6 can be understood by combining the slope changes observed in Model 4 and Model 5: i.e., transitioning gradually from the behavior of Model 5 to that of Model 4. In Model 6, the light curve up to approximately 2 days, resulting from the density profile within $0.2 \lesssim \beta \leq 0.35 $, resembles those of Model 5, but squeezed. However, as time progresses, the density profile of Model 6 exhibits a well-like dip, similar to that of Model 4, resulting in a change in the decay rate at around 3 days.

Since our work is to explore the change in kilonova emission patterns by varying the ejecta velocity profile, we do not narrow our focus to one or two ejecta components that have distinct properties such as ejecta velocity, timescale of ejection, and electron fraction $Y_e$ \citep[e.g.,][]{Kasen2017, Siegel2017, Radice2018}. In this work, we utilized ejecta speeds ranging from 0.08 to 0.35 c, which covers the speeds of dynamical and disk-wind components. The changes in minimum and maximum ejecta velocities can stretch or compress the lightcurve features. The duration of ejection, $\tau_d$, does not significantly impact the results because the timescale relevant to the observation, $t$, is sufficiently large (i.e., $t \gg \tau_d$).

However, we acknowledge that the location and duration of these variations in the kilonova lightcurve heavily depend on the element composition (or $Y_e$ of the ejecta). By varying the composition, we would expect to observe similar features in the lightcurve, but to have slight changes in the characteristic wavelengths and durations. 

In addition, temperature can play an important role in shaping the lightcurve behavior. In this work, we adopt a continuously decreasing temperature profile (see Equation~\ref{eq:temp}) to investigate the causality between the velocity profile and the resulting lightcurve features. However, the overall rate of kilonova luminosity decay is heavily contingent upon how the temperature of a heating source cools over time. This implies that a rapid cool or re-heating of the source could lead to obscure or amplify some lightcurve features caused by the irregular density profile.

Lastly, numerical-relativity simulations for the merger of binary compact objects \citep[e.g.,][]{Kyutoku2015, Radice2018} present the mass ejection rate as a function of time, $\dot{M}(\tau)$. We may obtain an intricate density profile in this case as well (like in Models 4 to 6), leading to similar behaviors in the kilonova lightcurve.

To date, only a few kilonovae have been observed, mainly because of their short lifetime and relatively low luminosity. However, with the enhancement of follow-up strategies using current telescopes and also with the development of advanced future telescopes like the 7-Dimensional Telescope \citep[7DT;][]{Im2021}, the chances of detecting kilonovae are expected to increase significantly \citep[e.g.,][]{Cowperthwaite2019, Chase2022, Ekanger2023}. In this circumstance, this work proposes a model to explain peculiar kilonova lightcurves that might be observed in the near future.

\acknowledgments
\section*{Acknowledgement}
We are deeply appreciative of the invaluable comments provided by an anonymous reviewer, which have greatly enhanced the quality of our work. This work was supported by the National Research Foundation of Korea (NRF) grant, No. 2021M3F7A1084525, funded by the Korea government (MSIT). This research made use of \textsc{tardis}, a community-developed software package for spectral synthesis in supernovae \citep{Kerzendorf2014, tardis}. The development of \textsc{tardis} received support from GitHub, the Google Summer of Code initiative, and from ESA's Summer of Code in Space program. \textsc{tardis} is a fiscally sponsored project of NumFOCUS. \textsc{tardis} makes extensive use of Astropy and Pyne.

\software{\textsc{tardis} (\url{https://zenodo.org/record/8128141})}

\bibliographystyle{aasjournal}
\bibliography{references}

\end{document}